# Negative differential spin conductance in doped zigzag graphene nanoribbons

Ting-Ting Wu,[1,2] Xue-Feng Wang,[1][*] Ming-Xing Zhai,[1,2] Hua Liu,[1,2] Liping Zhou,[1] and Yong-Jin Jiang[2]

[1]Department of Physics, Soochow University, Suzhou, China 215006

[2]Department of Physics, Zhejiang Normal University, Jinhua, China 231004

The spin dependent charge transport in zigzag graphene nanoribbons (ZGNRs) has been investigated by the nonequilibrium Green's function method combined with the density functional theory at the local spin density approximation. The current versus voltage curve shows distinguished behaviors for symmetric and asymmetric ZGNRs and the doping on the ZGNR edges can manipulate the spin transport. In special cases that a Be atom is substitutionally doped on one edge of the symmetric ZGNRs, one spin current shows negative differential resistance while the other increases monotonically with the bias. This property might be used to design spin oscillators or other devices for spintronics.

[*] Email: xf_wang1969@yahoo.com

As one of the most promising materials for electronic and spintronic devices, graphene nanoribbons (GNRs) have been attracting increasing interest.[1-4] Compared with carbon nanotubes, graphene not only has a simple structure, but also can be easily prepared and cut into GNRs along a certain direction in experiments. There are two types of GNRs: zigzag edge nanoribbons (ZGNRs) which are cut along the zigzag dimer chain and armchair edge nanoribbons (AGNRs) which are cut along the direction perpendicular to the zigzag dimer chain.[5,6] Furthermore, ZGNRs are found to be magnetic semiconductor with spin-polarized edge states, which are ferromagnetically ordered but anti-ferromagnetically coupled to each other. This suggests that ZGNRs may play an important role in quantum electronic devices in the future and have ample potential in spintronic applications.[7,8]

In general, we can manipulate the electronic properties of GNRs by defect,[9-11] impurity doping,[12-14] adsorption,[18] chemical functionalization,[19-21] external field,[4,22] and geometry etc. This provides the methods for applications in electronic and spintronic devices. Li *et al.*[23] have discussed the role of structure symmetry by assuming spin-unpolarized transport in ZGNRs. They observed that the transport depends strongly on the symmetry while both symmetric and asymmetric ZGNRs show metal behavior. H. Ren *et al.* have explored the realization of negative differential resistance or conductance (NDR) in armchair

GNRs using a no-spin model.[24] Furthermore, it has been realized recently that the spin can play an important role in GNR properties like opening energy gap near the Fermi energy and making ZGNRs semiconductors. Gorjizadeh *et al.*[7] have reported the spin and band-gap controlling by *ab initio* calculation in doped GNRs. Park *et al.*[14] have studied the influences of B or N impurities on GNR devices with Au electrodes. It has been observed that the spin states near the Fermi energy are responsible for the spin-polarized current. Y. Ren and Chen[11] have studied the effect of Stone-Wales defects on transport in ZGNRs and emphasized the effect of the structure symmetry.

In this paper, we present a spin dependent study on the electronic and transport properties of perfect and Be edge-doped ZGNRs. The study illustrates that Be-doping is a feasible way to realize the NDR for only one spin-component which has its unique advantage in spintronics. As we know, the substitutional doping of the alkaline earth metal Be atoms in Silicon is feasible.[15] The band structure of armchair GNRs doped with the alkaline earth metal Mg atoms has been calculated by N. Gorjizadeh *et al*[16] and our first principle calculation suggests that the substitutional doping of Be atoms in ZGNRs is stable. In the following, we use the 3-, 7-ZGNRs and the 4-, 8-ZGNRs to represent asymmetric ZGNRs and symmetric ZGNRs, respectively. The spin transmission spectrum and other electronic properties indicate a semiconductor behavior in perfect

ZGNRs. Near the Fermi energy, edge states of opposite spins in the ZGNRs are spatially separated but degenerate in energy. In the equilibrium at zero bias, the spin-up (down) electrons are localized near the lower (upper) edge. When a bias is applied in the infinite extension direction of ZGNRs, asymmetric ZGNRs act as ordinary semiconductors, whereas symmetric ZGNRs show negative differential resistance effects because of the appearance of localized electronic states in the existence of voltage drop.[11] In edge doped ZGNRs, the spin degeneracy is broken and the systems become spin polarized. In asymmetric ZGNRs the current versus voltage curve is similar to that of perfect ZGNRs but with a lower value. One peculiar phenomenon occurs in the Be doped symmetric ZGNRs where the negative differential conductance happens for only one spin while the current of the other spin increases monotonically with the bias voltage.

As illustrated in Fig.1, the atomistic systems used in the calculation are $N$-ZGNRs ($N$ zigzag chains in width) partitioned into three regions: the semi-infinite left electrode (L), the central scattering region (C), and the semi-infinite right electrode (R). The length of the central scattering region is chosen as ten unit cells to ensure that the screening effect of the impurity does not reach to the electrodes. The semi-infinite electrodes are described by couples of unit cells due to their periodicity. The electrodes and the central scattering region in this system are parts of the same

ZGNRs so the interfaces between them have less effect on the transport than those in conventional metal - conductor - metal systems.[17] In this work, the quantum transport is studied using the *ab initio* technique implemented in the ATK (Atomistix ToolKit) package which is based on the nonequilibrium Green's functions method and the density function theory.[25,26,27] The exchange-correlation potential resorts to the spin dependent local density approximation (LSDA). To obtain accurate enough result, we expand the wave function on a basis set of double zeta orbitals plus one polarization orbital (DZP). A 15 Å vacuum region is adopted to separate the ribbons in neighbor supercells far away enough and ensure the suppression of the coupling between them. The structures are optimized until the atomic forces are less than 0.02 eV/ Å. The C-C bond length is 1.42 Å, the C-Be one 1.67 Å, and the C-H one 1.39 Å. In the calculation, the energy cutoff is 150 Ry, the mesh grid in the *k*-space is $1\times1\times100$, and an electronic temperature of 300 K is used in the technique of the real-axis integration for the non-equilibrium Green's functions. The spin-dependent current $I_\sigma$ through the central region is evaluated by the Landauer-Büttiker formula:

$$I_\sigma = \frac{e}{h}\int_{-\infty}^{\infty} T_\sigma [f(E-\mu_R) - f(E-\mu_L)]dE.$$

Here $T_\sigma$ is the electronic transmission of spin $\sigma$, $f$ the Fermi-Dirac distribution function, and $\mu_i$ (i=L, R) the chemical potential of the left and right electrodes.

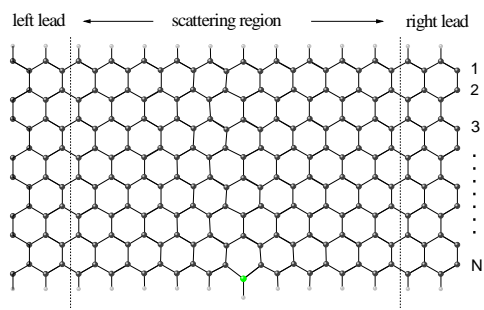

FIG. 1. Atomistic scheme of the two-probe system of an *N*-ZGNR with one Be atom (green dot) substitionally doped on the lower edge of the scattering region.

The spin-up (solid) and spin-down (dotted) current versus voltage (I-V) curves of the perfect ZGNRs are plotted in Fig. 2. All ZGNRs have semiconductor characteristics and the threshold voltage decreases with the ZGNR width. The spin effect is negligible for the I-V curves of the perfect ZGNRs. The transmission gap approximates to 0.36 eV, which is different from the zero gap prediction previously obtained by assuming no-spin polarization.[23] On the other hand, similar to the no-spin result, the symmetric and asymmetric ZGNRs show tremendously different transport characteristics, although they have similar electronic structures and transmission spectra at zero bias. When the width of the ZGNRs is odd, corresponding to asymmetric ZGNRs, the I-V curves become approximately linear at bias above the threshold voltages as illustrated in Fig. 2(a) and (c). The I-V curves of the symmetric ZGNRs, however, appear nonlinear with strong NDR effect as shown in Fig. 2(b) and (d).

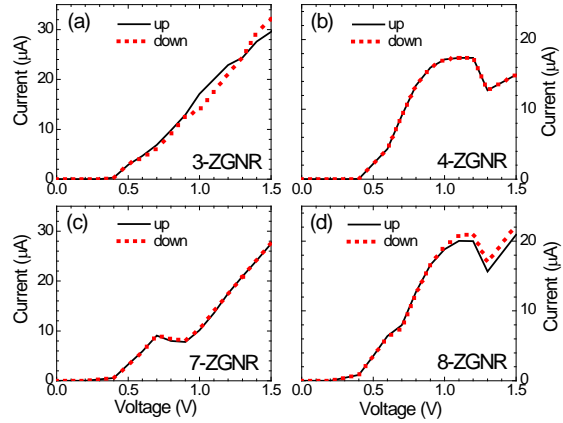

FIG. 2. Spin-up (solid) and spin-down (dotted) currents as functions of the bias voltage for perfect ZGNRs of different width: (a) 3-ZGNR, (b) 4-ZGNR, (c) 7-ZGNR, and (d) 8-ZGNR.

Usually the electronic properties are mainly determined by the states and electrons around the Fermi level. When the bias is below 0.36 V, the threshold voltage, the density of states in the transport window (the energy range between the left and right chemical energies) vanishes in one of the electrodes so that few electrons can pass through the central scattering region from the left to the right electrode and almost zero transmission or conductance is observed. As the bias goes up, the energy bands of the electrodes shift with the corresponding chemical energies and mismatch with each other. The transmission becomes finite in the middle of the transport window when the mismatch is near or larger than the energy gap and the electrons can tunnel from the valance band of one electrode to the conduction band of the other. Consequently, the current becomes finite at high bias. Similar to the previous observation,[11] the different characteristics of the I-V curves in asymmetric and symmetric ZGNRs result from the transmission spectrum of different structure symmetries.[23]

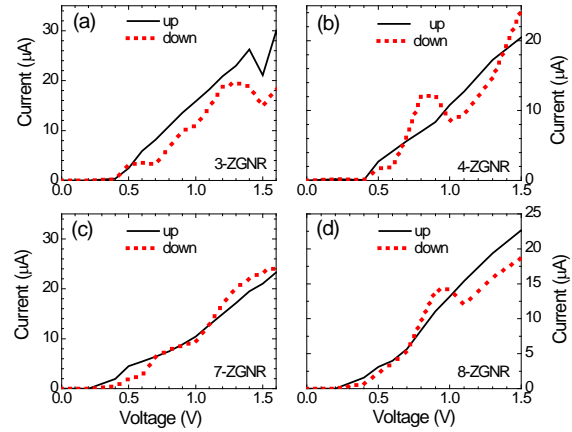

FIG. 3. Spin-up (solid) and spin-down (dotted) currents as functions of the bias voltage for the Be edge doped ZGNRs of different width: (a) 3-ZGNR, (b) 4-ZGNR, (c) 7-ZGNR, and (d) 8-ZGNR.

In Fig. 3, we present the I-V curves of the Be edge doped ZGNRs as schemed in Fig. 1, where a C atom is substituted by a Be atom on the lower edge of the central scattering region. The spin degeneracy in the system is obviously broken and the current becomes strongly spin dependent. The spin polarization defined by $\eta = (I_\uparrow - I_\downarrow) / (I_\uparrow + I_\downarrow)$ fluctuates between positive and negative values. The spin-polarization of 3-, 4-, 7-, and 8-ZGNR can achieve 43%, 37%, 63%, and 38%, respectively. In the studied cases, the asymmetric ZGNRs can reach to a higher spin-polarization ratio compared to the symmetric ones. In addition, we observe higher spin-polarization than the value previously reported (up to 39%) in the B atom doped ZGNRs connected to Au electrodes.[14] Overall the impurity atom reduces the current above the threshold bias except for symmetric ZGNRs at bias above 1.3V where the NDR occurs in perfect ZGNRs. Below the threshold bias, the impurity atom enhances the electronic transport especially for the spin-up channel. This effect softens the threshold edge and increases significantly the spin-up transmission for wider ZGNRs as indicated in Fig. 3(c) and (d). The most interesting effect in the Be edge doped ZGNRs is that in the symmetric ZGNRs only the spin-down I-V curve has a NDR region and it appears at relatively low bias (1 eV) while the spin-up current increases with the bias monotonically as shown in Fig. 3(b) and (d). Because the spin-down current is away from the spin-up current in real space, electric field domains may form only for the spin-down current due to the Gunn's effect. It is then possible to fabricate spin oscillators based on this spin Gunn's effect. The Mülliken analysis indicates that the Be impurity atom greatly weakens the atomic magnetism of the C atoms around it.

To understand better the NDR effect appeared in the ZGNRs, we present some details of the transmission spectrum and the spatial local density of states (LDOS). Above the threshold bias, the current increases with the bias because electrons in the valence band of one electrode can tunnel across the energy gap and the central region to the conduction band of the other electrode. It is then sensitive to the bridging states in the central scattering region, the coupling strength between the states in the central region and the electrodes, and the width of the transport window. The typical transmission spectra are like that shown in Fig. 4 (a), where the transmission spectrum of a perfect 8-ZGNR at the bias voltage 1.2 V (solid) and 1.3 (dotted) is plotted. When the bias increases from 1.2 V to 1.3V, the transport window widens from [−0.6, 0.6] V to [−0.65, -0.65] V and the effective nonzero transmission range widens from [-0.42, 0.46] V to [-0.47, 0.5] V. However the transmission decreases in the middle of the transport window, resulting in the decrease of the current and the NDR at bias 1.3V as illustrated in Fig. 2(d). The isosurface of spatial LDOS in the system at bias 1.2 and 1.3 V plotted in Fig. 4(b) and (c) indicates that the transmission reduces because the electrons in the central region become more localized.

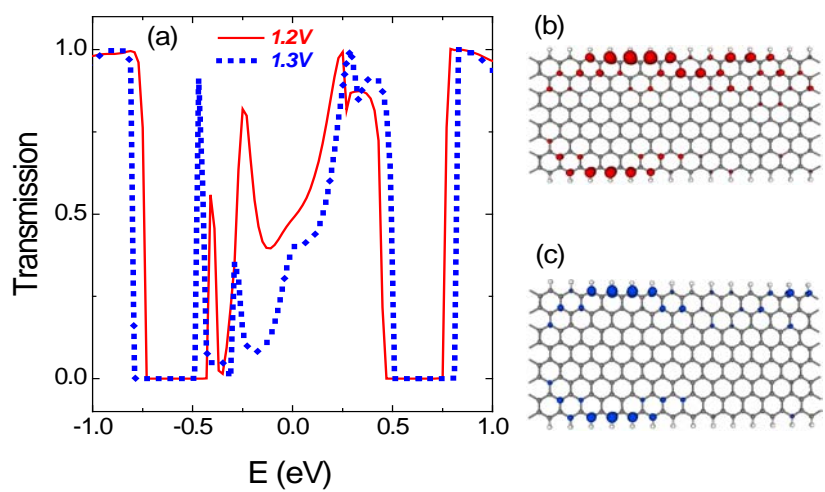

FIG. 4. (a) The transmission spectrum at the bias voltage 1.2 (solid) and 1.3 V (dotted) for the perfect 8-ZGNR. The corresponding isosurfaces of the spatial LDOS at value 0.032 Å$^{-3}$eV$^{-1}$ for electrons of energy −0.2 eV are shown at bias (b) 1.2 and (c) 1.3 V.

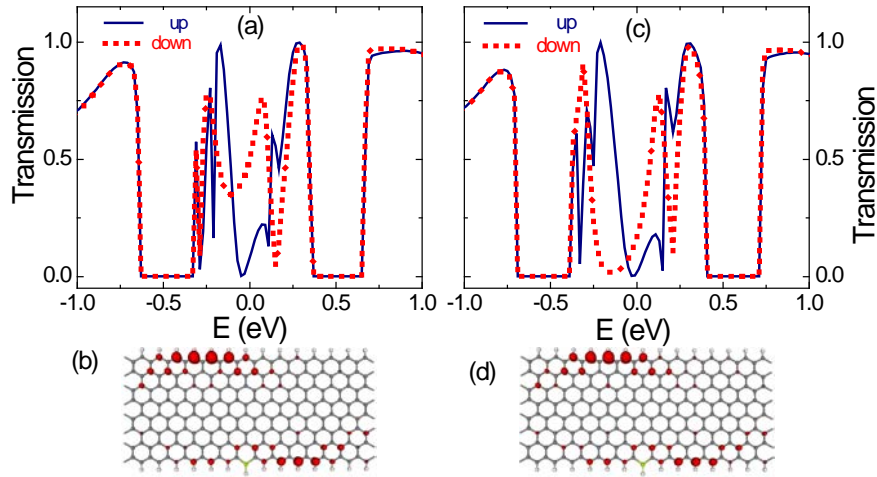

FIG. 5. The spin-up (solid) and spin-down (dotted) transmission spectrum at (a) 1.0 V and (c) 1.1 V for the edge doped 8-ZGNR. The corresponding isosurfaces of the spatial LDOS at value 0.032 Å$^{-3}$eV$^{-1}$ for spin-down electrons at energy E=0 are plotted in (c) and (d), respectively.

In Fig. 5 (a) and (c), we plot the transmission spectrum of the Be edge doped 8-ZGNR at bias voltage 1.0 and 1.1 V, respectively. The transmission is strongly spin dependent inside the transport window, [−($V_b$-0.36 V)/2, ($V_b$-0.36 V)/2] with $V_b$ the bias voltage, and this explains the spin dependence of current due to the symmetry breaking. The transmission profile of up-spin does not change much as the bias increases from 1 V to 1.1 V. The transmission of down-spin, on the contrary, decreases quickly and forms a transmission gap when the bridging states in the central region become more localized as indicated by the spatial LDOS illustrated in Fig. 5 (b) and (c). As a result, the spin-up I-V curve in the 8-ZGNR is monotonic while the spin-down one enters the NDR regime when the bias voltage reaches to 1.0 V. This phenomenon happens for symmetric ZGNRs indicating that the symmetry plays a crucial role in these devices. Our study shows that a single impurity atom in the edge of the ZGNRs can effectively manipulate the transport properties and control the occurrence of NDR in the ZGNRs for spintronic devices. It is worth noting that our setup is a modeling one and might not reflect the situation of corresponding real systems.

In summary, using the *ab initio* calculation, we have investigated the spin-dependent transport property of ZGNRs. The effects of the symmetry and impurity have also been considered. We find that ZGNRs are semiconductor with an energy gap of 0.36 eV due to the interaction between the spin-polarized edge states. The symmetric and asymmetric ZGNRs show different I-V properties. The I-V curves are almost linear in the asymmetric ZGNRs systems. The substitution of an edge C atom by a Be impurity atom suppresses the local magnetization around the atom and results in the spin polarization of current. In symmetric ZGNRs, the spin-up current has a monotonic I-V curve while the spin-down I-V curve shows a strong NDR effect, making the structure a promising material for spintronics.

This work was supported by the National Natural Science Foundation of China (Grant Nos. 11074182, 10804080, and 91121021) and Natural Science Foundation of ZheJiang Province, China (Grant No. 7080383). Part of the calculations was performed at the Center for Computational Science of CASHIPS.